# Quantum Computing at the Frontiers of Biological Sciences

Prashant S. Emani[1,2]*, Jonathan Warrell[1,2]*, Alan Anticevic[3], Stefan Bekiranov[4], Michael Gandal[5], Michael J. McConnell[4,6], Guillermo Sapiro[7], Alán Aspuru-Guzik[8,9,10,11], Justin Baker[12,13], Matteo Bastiani[14], Patrick McClure[15,16], John Murray[17,18], Stamatios N Sotiropoulos[14], Jacob Taylor[19,20], Geetha Senthil[21], Thomas Lehner[21#], Mark B. Gerstein[1,2,22,23#], Aram W. Harrow[24#]

* These authors contributed equally.
# Corresponding authors.

[1]Program in Computational Biology and Bioinformatics, Yale University, New Haven, CT 06520, USA.
[2]Department of Molecular Biophysics and Biochemistry, Yale University, New Haven, CT 06520, USA.
[3]Yale School of Medicine, Department of Psychiatry, New Haven, CT 06511, USA.
[4]University of Virginia School of Medicine, Department of Biochemistry and Molecular Genetics, Charlottesville, VA 22903, USA.
[5]Department of Psychiatry, Semel Institute, David Geffen School of Medicine, University of California–Los Angeles, 695 Charles E. Young Drive South, Los Angeles, CA 90095, USA.
[6]Univ. of Virginia School of Medicine, Department of Neuroscience, Charlottesville, VA 22903, USA.
[7]Department of Electrical and Computer Engineering, Duke University, Durham, NC 27708, USA.
[8]Department of Chemistry, University of Toronto, 80 St. George Street, Toronto, Ontario M5S 3H6, Canada.
[9]Canadian Institute for Advanced Research (CIFAR) Senior Fellow, Toronto, Ontario, M5S 1M1, Canada. [10]CIFAR Artificial Intelligence Research Chair, Vector Institute, Toronto, Ontario, M5S 1M1, Canada [11]Department of Computer Science, University of Toronto, 40 St. George Street, Toronto, Ontario, M5S 2E4, Canada.
[12]Schizophrenia and Bipolar Disorder Program, McLean Hospital, Belmont, MA 02478, USA.
[13]Department of Psychiatry, Harvard Medical School, Boston, MA 02114, USA.
[14]Sir Peter Mansfield Imaging Centre, School of Medicine, University of Nottingham, Nottingham, NG7 2RD, UK.
[15]Machine Learning Team, National Institute of Mental Health, Bethesda, MD 20892, USA.
[16]Section on Functional Imaging Methods, National Institute of Mental Health, Bethesda, MD 20892, USA.
[17]Department of Psychiatry, Yale University School of Medicine, New Haven, CT 06511, USA
[18]Department of Physics, Yale University, New Haven, CT 06511, USA.
[19]Joint Center for Quantum Information and Computer Science, University of Maryland, College Park, MD, 20742, USA.
[20]National Institute of Standards and Technology, Gaithersburg, MD, 20899, USA
[21]Office of Genomics Research Coordination, National Institute of Mental Health, Bethesda, MD 20892, USA.
[22]Department of Computer Science, Yale University, New Haven, CT 06520, USA.
[23]Department of Statistics and Data Science, Yale University, New Haven, CT 06520, USA
[24]Center for Theoretical Physics, Department of Physics, Massachusetts Institute of Technology, Cambridge, MA 02139, USA.

# ABSTRACT

The search for meaningful structure in biological data has relied on cutting-edge advances in computational technology and data science methods. However, challenges arise as we push the limits of scale and complexity in biological problems. Innovation in massively parallel, classical computing hardware and algorithms continues to address many of these challenges, but there is a need to simultaneously consider new paradigms to circumvent current barriers to processing speed. Accordingly, we articulate a view towards quantum computation and quantum information science, where algorithms have demonstrated potential polynomial and exponential computational speedups in certain applications, such as machine learning. The maturation of the field of quantum computing, in hardware and algorithm development, also coincides with the growth of several collaborative efforts to address questions across length and time scales, and scientific disciplines. We use this coincidence to explore the potential for quantum computing to aid in one such endeavor: the merging of insights from genetics, genomics, neuroimaging and behavioral phenotyping. By examining joint opportunities for computational innovation across fields, we highlight the need for a common language between biological data analysis and quantum computing. Ultimately, we consider current and future prospects for the employment of quantum computing algorithms in the biological sciences.

In an era of increasingly collaborative efforts towards unravelling the complexities of biology, one may, arguably, posit the existence of two broad epistemological tendencies: first, an approach towards greater clarity and depth in particular fields, whether relying on intensive technological, theoretical or computational development, that aims to comprehensively explore a specific aspect of a biological system; and second, a recognition of the need to knit together the disparate experimental and conceptual threads across the vast spectrum of length, time and system-size scales inherent in biology into a single, coherent framework. The field of X-ray crystallography is an unquestionable success story of the first epistemological approach: consistent improvement in beam-line technology and computational exploration of the immense space of candidate structures under experimental constraints, and detailed theories of biomolecular mechanisms came together in exquisitely characterized biological narratives on the structure of molecules. The achievements of the early phases of the Human Genome Project, directed at the construction of a high quality human reference haplotype through the intensive generation of genomic contigs to be stitched together, also collectively form an example of a sustained program with a clearly defined, focused end goal. Proponents of the second epistemological track have driven large-scale data collection efforts such as the UK Biobank[1] and the NIH's All of Us initiative[2], which aim to serve societal healthcare needs by integrating datasets ranging in scope from genomics to imaging to behavioral and disease phenotyping.

While no sharp delineation between the two tendencies exists, one could characterize them by the source of complexity that demands and drives innovation. In one, researchers must contend with the inherent challenge of answering well-defined questions, even where sharply targeted experimental assays exist. In the other, complexity arises from the lack of clear conceptual bridges between different scales and experimental domains, and the non-overlapping sets of assumptions among theories with different realms of applicability. For example, researchers are investing significant time and funding in surmounting the known technical issues of single-cell genomics so as to better quantify the mosaicism inherent in tissues. At the same time, it remains to be determined how the regional heterogeneity of these cells (determined in large part by genomics) couples together with differences in electrophysiological response (patch-clamp recordings) and neural system level connectivity patterns measured via functional magnetic resonance imaging (fMRI) to yield the complexity of mammalian brain function.

Addressing both these sources of complexity necessarily requires research-area-specific experimental and theoretical advances, but the concurrent growth in computational power also opens up the possibility of outsourcing some of the analytical burden to high-throughput computing resources. The significant interest in large-scale computing infrastructure shown both by governmental and private entities underscores the potential utility for the scientific community to explore new ways of interfacing with cutting-edge computing technologies. These include expansions of current super-computing and other massively parallel computing facilities, but also considerations of entirely new computing paradigms. In this piece, we specifically explore the potential of the emerging field of quantum computing in informing questions of significant complexity in biology. Recent technological developments have carried quantum computing capabilities from the realm of academic exploration to commercial opportunities[3–6]. While the scale is not currently competitive with classical technologies, there is substantial excitement in the eventual promise of this new field, and we hope to provide an entry point for members of the scientific community to certain aspects of the discussion surrounding quantum computing. This effort is especially timely given the recent passing of the U.S. National Quantum Initiative Act 2018[7], which calls for the implementation of a National Quantum Initiative directed towards the development of quantum information science and technology[8], as well as the European Quantum Technologies Flagship, a ten-year effort with the advancement of quantum computing technology as one of the primary targets[9], and the UK's National Quantum Technologies Programme[10].

In the interests of clearly identifying areas of possible gain in the use of the quantum paradigm, the NIMH convened a special virtual workshop in 2018 centered on addressing computational challenges in genomics and neuroscience via massively parallel and quantum computing[11]. The current article is a product of discussions initiated during that workshop. We present a focus towards the fields of genetics, genomics, neuroimaging, and deep behavioral phenotyping, especially as applied to the study of the human brain. We highlight these areas as they serve to exemplify the two aforementioned sources of complexity: separately, each field (and sub-field) presents an incredibly rich set of questions and problems that are in some cases already pushing the limits of classical computational capability; in combination, they represent a multi-scale challenge that starts at the molecular scale through the cellular and tissue levels, to brain architecture and, eventually, to complex human behaviors and its disorders. The study of the emergent properties of the human brain, such as cognition and behavior, is a uniquely challenging multi-level endeavor that demands pioneering approaches in computation.

Our approach in the following piece is to first present a primer on quantum computation to familiarize the reader with the basic concepts and language of quantum computing. We then proceed to an evaluation of some of the many challenges in genetics, functional genomics, neuroimaging and deep behavioral phenotyping, as well as in their integration. Finally, we discuss ways in which quantum algorithms that map onto those methodological issues may provide much needed improvements in computational efficiency.

**Classical versus Quantum Circuits: State of the Art and Opportunities for the Future**

Quantum computers (QCs) promise a new form of computing that would be qualitatively different from any previous ("classical") form of computation[12]. While QCs are technically more difficult to build, and the best current general-purpose quantum computers have only 50-100 qubits, they can solve some problems with a time that grows more slowly as a function of the input size. The term "qubit" refers to a quantum two-level system, such as the spin of a spin-½ particle. Qubits can be thought of as a generalization of classical bits (cbits) in that cbits can be in states 0 or 1, while the state of a single qubit is described by complex numbers $\alpha_0$ and $\alpha_1$ satisfying $|\alpha_0|^2 + |\alpha_1|^2 = 1$. The power of quantum computers comes from scaling. A system of *n* cbits can be in $2^n$ different states while the state of *n* qubits is described by a complex unit vector of length $2^n$. These vectors (also called wavevectors or wavefunctions) can be transformed by multiplying them by unitary matrices, and in many cases this can be done efficiently. For example, the wavevector can be Fourier transformed using $O(n^2)$ elementary quantum gates. However, not all transformations can be done efficiently. The laws of quantum measurement also limit the amount of information that can be extracted from a quantum state. A full measurement of the state yields outcome $x$ with probability $|\alpha_x|^2$, destroying the state in the process. Thus, even though describing the quantum state of n qubits requires an amount of information that scales exponentially with *n*, measurement can only extract *n* bits of information. Finding a way to benefit from the exponential state space of quantum computers despite this and other limitations is the central challenge of quantum algorithms.

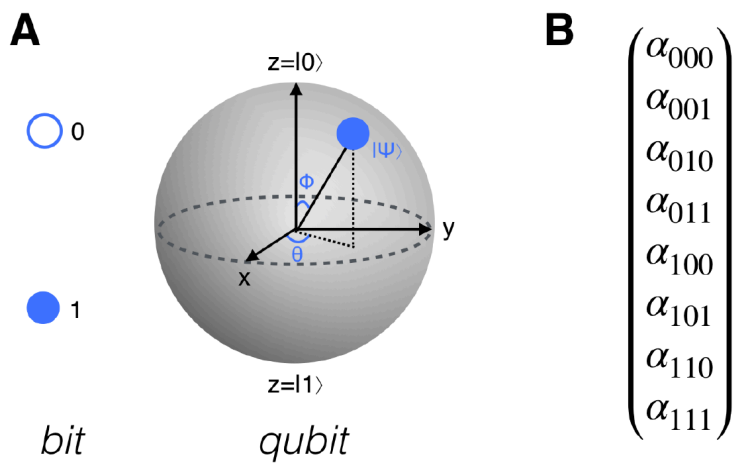

**Figure 1. A.** Conceptual illustration of bit vs. qubit. The state of a qubit can be represented by a point on the unit sphere with the North and South poles corresponding to the states 0 and 1 of a classical bit. **B.** The state space of 3 qubits is a $2^3$-dimensional complex vector.

Given the ubiquity of classical computers, the natural way to understand the strengths of quantum computers is by comparing their run-time scaling with the best-known classical algorithms. In some cases, these speedups are provably exponential: a QC with a few thousand error-corrected qubits could factor numbers that could not be factored using existing classical computers and algorithms in time less than the age of the universe (assuming some plausible conjectures about the classical computational complexity of factoring). In other cases, provable polynomial speedups are known: for example, given the ability to compute a function $f(x)$ where x takes on $N$ values, a QC can find the minimum value of $f(x)$ in only $O(\sqrt{N})$ evaluations of $f(x)$ while a classical computer would require $O(N)$ steps (assuming that f has no other structure we can exploit). On the other hand, for some problems, QCs are known to be no stronger than classical computers. And in many other cases, plausible heuristic algorithms have been proposed for QCs, whose performance is only incompletely understood. We discuss the different types of quantum algorithms in terms of their level of speedup over classical computers.

**Exponential speedup.** The main exponential speedups known are for cryptanalysis (dramatic but unlikely to be relevant here) and quantum simulation of molecules or other large quantum systems. If the properties of a molecule are not well captured by simple classical approximations then there is a good case to be made for using a quantum computer to make a better quality approximation computationally tractable. The advantage of a QC here arises from the exponentially growing dimension of quantum states. As a result, some promising cases for quantum advantage involve molecules with large numbers of active electrons, such as organometallic compounds[13]. Additionally, the recent claim to 'quantum supremacy' showed a significant speedup for the quantum computer relative to classical counterparts for the sampling of a random quantum circuit[6]. This is based on the plausible conjecture that classical computers need exponential time to simulate random quantum circuits.

**Polynomial speedup.** Typical polynomial speedups can be thought of as direct improvements of some classical algorithm. The most well-known of these is Grover's square-root search speedup[14], which is a quadratic improvement of classical brute-force search: given a search space of size N, brute-force search requires evaluating N points, while Grover search requires the equivalent of evaluating $O(\sqrt{N})$ points on a quantum computer. Other, more sophisticated, algorithms also admit provably quadratic improvements. For example, a classical algorithm might search over a tree of possibilities in a manner that can improve over brute-force search by sometimes being able to quickly prune entire subtrees. Such searches can also be quadratically improved quantumly, i.e. if the classical search process explores $N$ nodes than the quantum algorithm requires effort roughly equal to $\sqrt{N}$ times the effort to evaluate one node[15].

The strength of these algorithms is that they apply under extremely general conditions, such as needing to minimize an easily computable function. They also do not usually need more qubits than are already needed to compute the function. There is one important caveat about these algorithms. Suppose for concreteness that we are minimizing a function $f(x)$. Then a quantum computer would need to compute $f(x)$ in superposition over many different values of $x$, i.e. the computation could not leak any information about $x$ to any outside system. This would limit its ability to share the computation with a classical computer. Suppose, for example, that the evaluation of $f(x)$ were a memory- and time-intensive calculation for which quantum speedups were not known. Then using quantum computers to improve the minimization of $f$ would need to use qubits to perform this evaluation and could not offload the computation to a classical computer. This means that the overall speedup would be less than quadratic.

**Heuristic speedups.** Many of the most important algorithms for classical computers either lack formal proofs of correctness or are often run outside of the regime in which these proofs of correctness apply. These include Markov chain Monte Carlo (when rigorous upper bounds on mixing time are usually not known) and gradient descent applied to non-convex problems such as deep neural networks. For quantum computers, heuristic algorithms include adiabatic optimization[16] and the quantum approximate optimization algorithm (QAOA)[17]. The level of speedup provided by these algorithms over classical algorithms is in general unknown, and may be anywhere from an exponential improvement to no speedup. It is expected that as quantum computers are built, our understanding of the performance of these heuristics will improve, just as much of our understanding of the performance of classical heuristics comes from empirical evidence and not only theory. It should be noted that the same caveat about evaluating $f$ in superposition applies to most quantum heuristics as well.

**The source of quantum speedup.** One way to explain the power of quantum computers is by comparing them with randomized computers. The state of an n-bit classical computer using randomness could be viewed as a probability vector of length $2^n$, analogous to the amplitude vector of length $2^n$ that describes the state of a quantum computer. In each case, the entire vector cannot be read out from one run of the computer and the final output will only be a sample from a (hopefully useful) distribution. However, in a randomized computation, probabilities are always nonnegative, so they combine to a final answer simply by addition, while in a quantum computer, amplitudes are complex numbers. This means that different paths through a computation can "interfere" either constructively (if they have amplitudes with nearly the same phase) or destructively (if they have amplitudes with very different phases), analogous to the way that light and other waves can exhibit interference. While we often do not know how to take advantage of the rich possibilities offered by quantum interference, in some cases we can use them to achieve asymptotic speedups. Algorithms like Grover's are simple examples of this, making use largely of the fact that probabilities are obtained by taking the square of quantum amplitudes, so that a subroutine with a small success probability $p$ needs to be repeated only $O(1/\sqrt{p})$ times instead of $O(1/p)$ times[18]. The quantum Fourier transform (used in period finding and Shor's factoring algorithm) is a more sophisticated example of how complex-weighted transitions can be useful, and in some cases this can give rise to exponential speedups. On the other hand, some problems are known to not admit any quantum speedup, e.g. taking the parity of $N$ numbers requires time $O(N)$ on either a quantum or classical computer[19]. It is a major open research problem to determine when quantum speedup does or does not exist, and it is unlikely to ever be fully resolved, just as there is still no single theorem describing which problems can be solved by efficient classical algorithms.

**Opportunities for the future.** Existing quantum algorithms, for example, function minimization, are often written in terms of abstract and highly general functions. If biological applications can help motivate specific, mathematically well-posed tasks, then it may be the case that targeted quantum algorithm development can lead to improvement. This goal forms the core of this article and is discussed at length in the following section in the context of the study of the human brain. Here we briefly introduce some of the key areas of ongoing research in quantum computing, related to and providing the context for applications in biology.

*Machine learning and big data.* An important limitation of the models of quantum computers currently under development is that they cannot access large classical datasets in superposition (a similar limitation was discussed in the above paragraph on "Polynomial Speedups"). This means that they may be able to speed up complicated calculations on small datasets (e.g. finding the best Bayesian network) but have less advantage in solving problems on large datasets. One way to address this is with filtering or data reduction techniques, which select a small but hopefully representative sample of the data and use that as input to the optimization problem. Or the quantum computer could be used for "small data" problems where the difficulty comes from the complexity of the analysis. A more speculative possibility is a quantum hardware solution known as a qRAM (quantum RAM)[12,20], which would give a quantum computer the ability to coherently query a large classical dataset as a superposition of qubits. In other words, a superposition of input memory addresses would yield an output consisting of a superposition of memory cell contents. A qRAM would enable powerful quantum algorithmic primitives[20] but there are no proposals for scalable error-corrected qRAM, and it is not clear if it would ultimately be easier than making a large quantum computer[21].

*Simulation of classical and quantum systems.* Quantum algorithms have been developed for solving large linear systems of equations as well as Partial Differential Equations (PDEs) and boundary-value problems. These algorithms have excellent scaling with the size of the problem but not with other parameters, such as condition number or nonlinearity[22,23]. Along with other restrictions on the algorithms, this makes it a research question to determine how much speedup they can offer for specific applications.

There have been successful demonstrations of the application of quantum computation to problems in chemistry. A Variational Quantum Eigensolver (VQE) approach was used[24] to estimate the ground state energies of small molecules as a function of their component atomic separations. Briefly, short quantum circuits define a variational ansatz of trial solutions for the ground state and the circuit parameters are varied to minimize the energy using algorithms such as gradient descent. While the complexity of simulating quantum dynamics on quantum computers is well understood and is usually tractable, the success of VQE will depend on the quality of the ansatz and is an active area of ongoing research.

There has also been considerable interest in extending QC to biomolecular[25] and biological problems. A quantum annealing (QA) approach was employed in the exploration of the coarse-grained folding landscape of a six-amino acid peptide, within a 2D lattice framework[26]. In spite of the simplifying assumptions, the work serves to effectively demonstrate the scope for QA in searching large combinatorial spaces, as are ubiquitous in the field of protein folding simulations. QA was also evaluated against a set of classical methods on an optimization problem involving the search for the consensus DNA sequence motif of transcription factor binding[27]. In this instance, the authors trained a classifier (sequence is binding or non-binding) and a ranking algorithm (ranking sequences by binding affinity), finding a slight improvement of QA over classical approaches in the classification problem, and similar performance for the ranking task.

Quantum simulation of chemical reactions is known in principle to be possible on a quantum computer and as the practical details are fleshed out, this is expected to be an important application of quantum computers for applications both inside and outside of biology. One particular strength is in modeling dynamics, and there is evidence that energy transport (such as in photosynthetic complexes[28–32]) and electron transport (such as at redox sites of metalloproteins[33]) in biological molecules involves quantum effects that could potentially be more accurately modeled by a quantum simulation[34].

## Potential applications for Quantum Computing in Neuroscience

A unified model that leads from a molecular-genetic level understanding of the brain, through cell-based and regional analyses, to the interface with global structural/functional networks requires a whole host of methodologies applied in combination (for example, see ref.[35]). This diversity and complexity is reflective both of the ever-expanding tool set of experimental assays and of the varied complexity of measurement at different scales of analysis. We provide a minimal, but representative, sample of the challenges associated with these complex datasets by tracing a path from genetics and genomics, through neuroimaging, to behavioral phenotyping. We acknowledge that there are many other tools and levels of inquiry, but the focused choices here reflected the practical goal of illustrating the *computational* problem. We end the section with a view towards the integration of the results from each of these disciplines, in the hope of bridging the disparate scales and theoretical frameworks. We place special emphasis on tasks that are processing-intensive and that have recently been the focus of machine learning approaches. This set of tasks is juxtaposed with analogous quantum algorithms that could potentially be exploited to address some of these challenges in the future. Thus, we explore plausible quantum computing (QC) solutions for neuroscientific problems and posit open questions for eventual development of new computational paradigms. Where relevant, we will highlight the difference between conventional QC approaches and those requiring qRAM (as described in the previous section), as this difference has implications for near-term implementation of these methods.

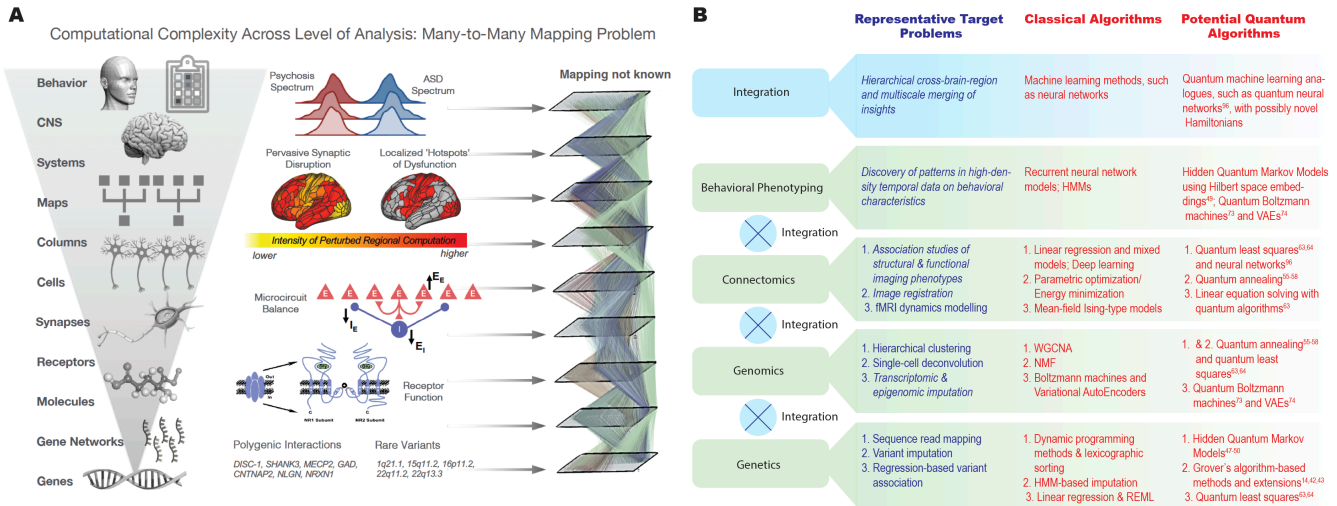

**Figure 2. A.** Illustration of the vast complexity when attempting the link levels of analyses from genomics to human behavior. This challenge remains in part due to interrogating the enormous search space for determining the mapping across levels, which constitutes a many-to-many probabilistic problem. Computational innovation will be a key effort to help close these gaps. Figure adapted with permission from ref. [36]. **B.** Schematic of the content of this article, highlighting some of the general problems within each of the scientific domains, the classical computing solutions, and the potential quantum computing solutions. The italicized items under the "Representative Target Problems" column represent problems that are especially challenging for current classical methods.

## *Genetics and sequence analysis*

An essential initial step in genetics and genomics is the matching of sequences of nucleotides and amino acids to reference databases, and, more specifically, the mapping of sequencing reads from a variety of experimental assays to genomes. The ubiquity of the sequence-matching and mapping processes, combined with the memory- and time-intensive computational needs, make these conceptual problems worthwhile targets of quantum computing improvements. Due to the incredibly large search space for a match to each sequence, any solution needs to contend with both memory (to hold a representation of the reference database or genome and information on the mapping) and speed concerns. Dynamic programming methods, such as the Needleman-Wunsch[37] and Smith-Waterman[38] algorithms enable queries of sequence strings against immense databases. These methods can potentially be reformulated as analogues of the Viterbi algorithm used in Hidden Markov Models (HMMs). While these methods have been mostly supplanted by the approximate but faster k-mer-based BLAST algorithm[39], an improvement in the general efficiency of dynamic programming methods could potentially allow optimal string alignment.

In the case of genomic read mapping, elegant classical algorithms were designed in response to the memory and speed challenges, such as the exploitation of the Burrows-Wheeler transform to efficiently perform DNA sequence alignments[40], and the use of seed-based approaches[41] to contend with mapping RNA reads to the boundaries between exons separated by large genomic distances. Both these methods are based on the construction of lexicographically sorted suffixes constructed from an immense reference genome, followed by scanning for matches to subsequences of the query read. If Grover's algorithm-based improvements in string matching search speeds could be exploited (see ref. [42] for $O(\sqrt{n} + \sqrt{m})$ speed-up for a n-length reference and m-length query), it may be possible to make the mapping process significantly faster. In fact, recent work has demonstrated the potential for even further speed gains under the assumption of unique membership of a query string within a reference database[43]. However, given the need for storing a large reference database (of the whole genome or exome), the lack of qRAM could limit any gains. Using the reads mapped to the reference genome, it is then possible to identify individual-specific sets of mutations, including single-nucleotide polymorphisms (SNPs). Subsequently, based on the shared sets of haplotypes across subpopulations, this first set of SNPs can be expanded by imputation of additional SNPs that co-occur with the original set with high probability. This imputation usually involves a maximization of a likelihood function achieved through the definition of a HMM[44,45]. Speed-ups in the genotype imputation problem are achieved through a variety of numerical approximations and optimized search strategies.

As can be seen, the set of prominent read-mapping and imputation algorithms share a mostly common mathematical language, arising from the fact that the genome consists of a linear sequence of nucleotides. The classical complexity of read-mapping problems varies depending on whether exact or inexact matches (including gaps) are considered ($O(n+m)$ and $O(nm)$ respectively, where n and m are respectively the reference genome/sequence and read lengths) and whether the reads are mapped independently or jointly (the latter being called multi-read alignment, with complexity $O(n^m)$, with m the number of reads and n the number of positions or transcripts they can be assigned to[46]). The size of the strings involved is such that a reduction in complexity of even the simpler mapping problems would be highly beneficial, although the string sizes also create potential problems for full-scale QC-based algorithms. The recent development of Hidden Quantum Markov Models (HQMMs)[47–50] opens the possibility of both simulating classical HMMs on conventional quantum circuits[49], as well as expanding model space beyond classical HMMs[47]. However, some of the more near-term approaches to seeking a quantum advantage could involve hybrid approaches, where classical circuits are employed for some modules of an algorithm, and quantum circuits would be used for certain optimizations: thus, the iteration through hyperparameter space in HMMs could be classical, with a quantum optimization of the maximal trajectory through state space. If data could be accessed in superposition (say, with a qRAM) then one could also replace the scanning process in suffix-array-based methods with Grover's algorithm, leading to quadratic speedups of this portion of read-mapping.

Another important category of genetic analyses is the construction of optimal trees that describe the relative proximity of genetic sequences. These could include: the generation of ancestral recombination graphs[51–53], that try to reconstruct historical relationships between the genomes of individuals while also allowing for segments of the chromosomes to have undergone recombination between parents; the construction of pathogen evolutionary trees in epidemiological studies based on their mutation patterns; the evaluation of mutational "mosaicism" among the cells in a single tumor, that may have a bearing on the medical response and virulence of the cancer. Tree reconstruction algorithms are designed to solve the optimization problem of simultaneously matching all the distance constraints between the individual genome segments. Many approaches exist, mainly involving sampling from the overall space of possible genealogies, with heuristics and simplifications including the treatment of the problem as approximately Markovian[54]. The search space S of coalescent trees underlying an ancestry analysis is massive ($|S| = \frac{n!(n-1)!}{2^{n-1}}$, where n is the number of individuals/sequences in the analysis), and the search space for ARGs is exponentially larger, since an ARG assigns a coalescent tree to every base in a sequence (hence an exhaustive optimization is $O(|S|^m)$, with m the length of the sequences). Potentially, the small input size (nm) and massive search space make this a candidate open problem for quantum speed-up (without using qRAM) using well-studied optimization methods such as quantum annealing (QA)[55–58].

After uncovering the genealogical structure in samples, the SNPs from genotyping studies can then be statistically linked to observable phenotypes directly associated with psychiatric or neurodegenerative disorders (GWAS) or to quantifiable "endophenotypes" or "quantitative traits" (cell/tissue gene expression levels, methylation, epigenetic markers, cell fractions, etc.) putatively involved in the disruption of gene regulatory networks. The linkage is aided by the removal of potential environmental or technical confounding factors, say, through Bayesian inference analysis[59]. Large-scale GWAS and quantitative trait loci (QTL) analyses are problematic for near-term quantum approaches, since they typically involve the analysis of large datasets. The subsequent evaluation of the total SNP heritability often proceeds through the avenue of linear mixed models, and the associated genetic variance estimations are carried out through techniques such as the restricted maximum likelihood (REML) method (for example, as implemented in GCTA[60]). Similar to GWAS analyses, the efficiency of these methods suffers from the need to carry out operations on large matrices, such as the genetic relatedness matrix (GRM), to solve systems of linear equations. The exploration of alternative classical methodologies is an area of active research, yielding several promising results[61,62]. If qRAM approaches can be developed though, algorithms such as Quantum Least Squares[63,64] offer up to exponential speed-ups in such analyses through the ability to perform fast linear-algebra operations, although it is unclear how much advantage would remain after accounting for the time cost of the qRAM. However, if the dimensionality of the underlying linear regression problem can be reduced, there is potential for near-term conventional quantum annealers (such as that implemented by D-Wave[27]) to tackle these tasks as well.

The notion of SNP prioritization discussed above can also be extended to somatic variants. For example, single cell genome sequencing studies in the brain indicate that every neuron is likely to contain private somatic variants. While single nucleotide variants are especially common, as many as 30% of neurons harbor large structural variants that alter allelic diversity for dozens of genes. Incorporation of the mosaic genetic architecture of the brain with cells and circuits is a major challenge facing genetics and functional genomics. One possible analytical approach (Schizophrenia Genetics and Brain Mosaicism project[65]) is to identify single nucleotide and structural variants associated with the occurrence of psychiatric disorders, using machine-learning classifiers trained on case/control datasets. However, given the potentially large-dimensional parameter search space for the classification problem, classical computation could run into search efficiency issues. These issues could possibly be ameliorated with the aid of quantum machine learning methods[66], discussed in more detail in the following sections.

*Functional Genomics*

The causal chain by which genetic variation leads to expression in higher-level behaviors such as cognitive traits involves multiple levels of intermediate molecular-to-cellular-to-system-level steps, governed by complex developmental processes and gene-environment interactions. In essence, this is a probabilistic and dynamic many-to-many mapping problem, which may vary dramatically across the human population in relation to individual differences. Despite this complexity, a range of studies have shown that genetic risk for particular traits can be partitioned across 'intermediate' phenotypes, such as gene expression or chromatin binding profiles, leading to insights into disease etiology[67–69]; a direct approach to such analysis is to impute intermediate molecular phenotypes first, and use the imputed phenotypes to predict high-level traits[70]. However, intermediate molecular phenotypes are typically high dimensional, such as bulk transcriptome expression profiles in a particular brain region (~22K dimensional), and highly interdependent, meaning that simplifying assumptions of independence are often necessary. Possible models which can learn joint probability distributions over such levels of analyses include Bayesian Networks, undirected models such as Boltzmann Machines[71], and recent deep-learning approaches such as Variational Autoencoder (VAEs). Exact optimization of such models however is intractable: structure learning in Bayesian Networks requires optimization over a search space of all directed acyclic graphs, which is super-exponential ($O(n!\, 2^{n!/(2!(n-2)!)})$, where n is the dimensionality[72]). On the other hand, inference in Boltzmann machines requires a search over $O(2^n)$ states after binarization to calculate a gradient, and training VAEs requires the optimization of a non-convex objective function. Such problems may be potential candidates for quantum approaches: for smaller input sizes, approaches without qRAM may be developed to perform exact searches across the space of Bayesian networks, while approximate quantum analogues of Boltzmann machines and VAEs have been tested in simulation and experimentally[73,74]; empirical evidence suggests that these are able to draw on the possibility of tunneling between low-energy states during quantum annealing to perform more efficient optimization of these models. We note also that for all these models, prior knowledge of possible molecular interactions may be used during training to suggest causal interpretation of the networks learnt.

Considering the brain in particular, the human cerebral cortex is comprised of over 80 billion neurons each with unique connections to other neurons and support from non-neuronal cells. Neuronal diversity is vast. Broad categories of excitatory, inhibitory, and neuro-modulatory neurons are then subcategorized by distinct marker gene expression and neurotransmitters produced. Single cell transcriptomic approaches further refine these subcategories and identify unique subtypes of human neurons. This diversity colludes with the combinatorial behavior of gene-gene and gene-regulatory element interactions to yield a pattern of intra- and inter-regional variation that seems essential to the function of the brain. To capture this level of complexity, in addition to molecular phenotypes, intermediate phenotypes may be derived at the level of sets of genes (such as functional pathways), and cell-type proportions. Phenotypes at these levels can be derived through analysis of transcriptome and other molecular data; for instance, Weighted Gene Correlation Network Analysis (WGCNA) performs a version of hierarchical clustering to derive co-expression modules, which are enriched in gene pathways[75], and non-negative matrix factorization (NMF) based on 'marker-gene' profiles can be used to decompose bulk transcriptome data into components corresponding to cell-type fractions[71]. Exact optimization of these models is again intractable, where exact hierarchical clustering would require a search over |S| trees (with S as in the previous subsection), and NMF is a non-convex optimization problem[76]. The former may be a candidate for an exact quantum solution for small-scale problems, while both may benefit from approximate quantum annealing approaches (an annealing-based approach to NMF is found in ref. [77], which involves discretization of the weight-space).

*Mapping Neuro-Behavioral Variation in Humans via Neuroimaging and Deep Phenotyping*

The overarching scientific goal of so-called 'convergent' neuroscience is ultimately to link noted cellular-level mechanisms to system-level observations and ultimately behavioral variation. Rapidly evolving human multi-modal neuroimaging provides rich sources of high-dimensional information that can link measures at the level of brain areas and systems with the mechanisms underlying human behavioral variation. However, it is important to offer a sobering reminder: in the last decade a typical single fMRI voxel placed in human cortex of ~3.8 mm cubic in size contains ~5.5 million neurons, 2.2–5.5 × $10^{10}$ synapses, 22 km of dendrites and 220 km of axons. Today, that size is closer to 2 mm cubic, but gaps between levels of analysis still remain staggering[78]. The goal of mapping healthy human brain function is to enable the characterization of associated computations, which could be in turn linked to alterations in disease. Closing these explanatory gaps is paramount. On one level, we need massive population-level data to characterize genetic trends and also to quantify perhaps subtle, yet significant signatures of individual behavioral variability[1]. On a vastly different level, we need computational tools to link neuroimaging measures to cellular-level and molecular diversity and, in turn, back to population-level variation both within and across individuals. Moreover, there is a particular need for developing quantitatively and neurobiologically grounded neuroimaging phenotypes that could have clinical impact. This is especially acute given the current status of mental health diagnosis being primarily limited to DSM IV/ICD 10, which are mostly categorical. In this respect, we badly need novel methods for rapid and scalable dimension-reduction when analyzing big data and/or smart parallelization of otherwise massively serial computational tasks (e.g. massively univariate processing of single voxels in 10,000s of imaging datasets). In addition, building models of brain dynamics and structural changes across multiple scales is a key sub-problem in identifying disease variation, which can in turn lead to more precise patient segmentation. Quantitative techniques that are able to handle such high-dimensional data challenges across the neuro-behavioral ontological gap will be key to help map and inform variation across health and disease. Currently, the field of neuroimaging is continuously facing computational bottlenecks that require creative algorithmic solutions.

One of the core challenges in the realm of brain imaging is the accurate registration of query brain surfaces and volumes to reference images. For instance, one methodology available in the popular imaging processing platform FreeSurfer[79,80] employs a sequence of registration steps to first align the pial and cortical surfaces to the reference, and then discovers a volumetric transformation that carries points within reference image to match the query image[81]. The volumetric alignment procedure involves two stages: an initialization stage consisting of relaxing an elastic model of the image constrained by a match to the surface registration; an intensity-based optimization stage where voxel intensities are matched between images. Both of these stages involve the minimization of an energy functional over the transformation field. If the corresponding Hamiltonian can be mapped to an Ising-type model, the advantages of a quantum annealing approach could be brought to bear.

The aforementioned multiscale framework leading from molecular to behavioral characteristics has often been approached in terms of association studies as a first approximation to the full complexity: the association could be from genetic variants to imaging phenotypes, or from imaging-based phenotypes to behavioral and clinically relevant traits. The former is an instance of a GWAS, and the construction of large imaging databases such as that of the UK Biobank has enabled GWAS of structural and functional imaging phenotypes[82]. The underlying linear regression and mixed models can be targeted by quantum computing, as described in a previous section on GWAS. The latter type of association analyses can be identified as imaging-based biomarker identification studies. For instance, the presence of active psychotic symptoms in previously unseen individuals diagnosed with schizophrenia and bipolar illness can be predicted using dynamic connectome features derived from fMRI[83], and other approaches have shown that combining static and dynamic features from fMRI can predict a range of phenotypes[84]. Quantum analogues of these approaches (such as HQMMs[47–50]) may help train such models more efficiently.

Recently, computational neuroscience has been effectively used to inform and constrain human neuroimaging observations. A broad class of dynamical neural models may operate at the local circuit or global level, and use parameterizations based on known constraints (e.g. biophysical parameters) or learned *de novo*. Local and global neural dynamics are typically highly non-linear, producing difficult optimization problems in the case of parametric model fitting[85–87], and requiring a rich model-class for *de novo* learning methods. Classical models can relate the structure of such networks to features of their equilibrium distribution (or resting state): for instance, Ising models and second-order mean-field regional models to model resting-state fMRI observations[88–90]. Fluctuations at equilibrium exhibit complex interdependencies, and additionally the relationship between structure/genetics and the equilibrium connectivity state is, in general, highly non-linear, and only partially captured by available models. To probe the structure/genetic relation to connectivity, it may be feasible to exploit the conceptual overlap of gene regulatory and resting-state networks, and thus extend published Deep Boltzmann machines for modeling gene regulatory network equilibrium states[71]. In the quantum computing domain, models such as the Quantum Boltzmann machine (QBM)[73] and Quantum VAE[74], as discussed in the previous subsection, may be naturally applied to model such complex distributions, either to potentially improve the optimization of their classical analogues, or learn intrinsic quantum representations by optimizing transverse couplings in their quantum Hamiltonians[73]. Such models may be used to study the effects of genetic/structural variation on network properties[82,86], or impute molecular data when it cannot be observed directly (e.g. in living subjects). Further, quantum algorithms have been developed which have the potential to offer exponential speed-ups in the solution of linear differential equations[22,91]. Differential equation-based models of global brain dynamics have been proposed, which represent regional firing rates using a mean-field approximation[89,92]. These models can be fitted to fMRI functional connectivity data, by linearizing the initial stochastic nonlinear system of differential equations around a fixed point using the method of moments[89], and using methods such as Approximate Bayesian Computation to fit parameters[85]. Quantum linear system solvers[63,64] have the potential to better fit such models to data by increasing the efficiency with which parameters can be tested, and increasing the resolution.

Further, general purpose quantum solvers for nonlinear systems of differential equations have been proposed[23], although currently these seem unlikely to offer speed-ups over classical methods. Efficient general purpose solvers would eliminate the need for linear approximations, and allow more accurate fitting of neural dynamical models, particularly out of steady state (for example, transitions between resting-state and task-based fMRI), and this application may help motivate finding better quantum algorithms for nonlinear differential equations.

Ultimately, leveraging neuroimaging measures, the goal of human neuroscience is to build integrative 'multi-level' models that can connect underlying brain states to cellular phenomena and to observed behavioral patterns. This computational challenge is also particularly acute in the case of 'deep' behavioral phenotyping (e.g. digital 'real time' measures), which can generate massive amounts of continuously measured dynamical behavioral variables. In this situation, there is clear potential for 'very deep' optimization and the opportunity for massive state-space exploration. Relevant use-case scenarios include 'in-the-moment' clinical decisions that may require rapid computation. This also becomes directly applicable for longitudinal data collection dealing with real-time digital phenotype/mobile technology, which faces the challenge of rapid and precise data reduction. Whereas conventional behavioral measurements in humans have typically required a relatively small number of data points (e.g., fewer than 1000 total data points associated with responses on a variety of self-report measures, psychological tests, or tasks to probe for latent cognitive function), the rise of temporally dense sampling methods in humans, along with the potential to generate 'just-in-time' interventions on the basis of those signals, presents new computational challenges in the realm of behavioral measurements. For instance, rich, phenotypic characterization using high-resolution video and audio can be highly disease relevant in behavioral illness, and yet are rarely collected since they are identifiable in raw form and present operational challenges to data reduction and protection of participant privacy. Quantum computing may provide more efficient ways to encrypt and decrypt large files, and to generate privacy-preserving behavioral metrics at the highest temporal resolution to facilitate meaningful multi-level modeling of behavior. For extended, continuous recordings of multi-sensor data from mobile devices (e.g., accelerometer, GPS), robust identification of anomalous behavioral signals requires detection of subtle patterns (present across multiple, noisy sensor streams), which must be optimized by learning from signals accumulated both from the individual over time and from relevant population-level information. The complexity of these data presents an obstacle for learning algorithms, which have to deal with extremely high-dimensionality of data needed to informatively link non-linear dynamics of brain states (e.g. fMRI) and the influence of time dependencies relevant to behavioral mapping. Recent deep learning approaches using interpretable recurrent networks have provided a powerful means of learning such brain-state/behavior associations de novo by jointly modeling fMRI data and behavioral data[93]. Quantum analogs of neural network frameworks (such as QBMs[73] and QVAEs[74]) have the potential to discover novel structure in these datasets. Models such as Hidden Quantum Markov Models using Hilbert Space embeddings[49], provide alternative dynamical models with intrinsically quantum representations, which have been shown to have comparable or possibly improved performance relative to classical methods on small-scale problems through classical simulations. Further, there is evidence that quantum dynamical models such as HQMMs allow complex dynamics to be modelled with a reduced state-space[47,50] compared to classical models, albeit so far in toy models. These hidden-state methods could possibly be applied to the evaluation of the dynamics and the switching between underlying brain states (resting-state or task-based[94]).

*Integration across disciplines*

Stitching together insights across the aforementioned sub-fields, to yield a holistic picture of brain function, is an ongoing challenge. Just as classical mechanics can be seen to emerge from quantum physics at the mesoscale, different views of functionality emerge in the brain (and other biological systems) at different scales, which may be broadly divided into David Marr's tripartite classification of computational task (behavior), algorithmic (circuit dynamics) and implementational (molecular/cellular) levels[95]. While the extent to which quantum processes are relevant across different levels of Marr's hierarchy is unclear (see *Epilogue*), Quantum Machine Learning may help elucidate the interdependencies between levels through its ability to learn and simulate non-linear models, which are classically intractable. One of the more promising avenues involves mechanism-agnostic machine learning methods like deep neural networks, where biological insights are gained by interpreting the model retroactively. Such an interpretable framework would involve connections between modules such as gene regulatory networks on the one hand, and structural/functional neuroimaging parameters (e.g. cortical thickness, white matter integrity, dynamic functional connectivity, etc.) on the other. The exact nature of these connections could be altered in competing hypotheses: one could imagine a hierarchical network with the molecular phenotypes at the base, the emergent system-level phenotypes (neuroimaging-based) at a higher layer, and the behavioral phenotypes serving as prediction targets. An alternative framework would treat the molecular and neural system-level components as parallel factors in determining behavior, the neural system-level components having been influenced at an earlier, developmental stage, and not directly emerging from the molecular phenotypes per se but rather operating in dependent 'lock-step'. In this way, different architectures of relationships between levels of analysis may be constructed. In fact, the NIMH has recently supported efforts at building such multi-scale, "convergent neuroscience" approaches (https://grants.nih.gov/grants/guide/pa-files/par-17-176.html). Such an analysis could be aided by quantum neural networks (QNNs)[96] and quantum variational classifiers[97], designed for use on non-qRAM, gate-based quantum computers. Quantum variational classifiers have been shown to be able to successfully classify states that were designed to be hard to simulate classically[97]. This hints at the greater generality of such circuits than their classical counterparts. Furthermore, insight from physical, mechanistic models may also be brought to bear on the integration of genomics with neural system-level mechanisms and ultimately human behavior.

**Epilogue**

While the field of QC is currently experiencing great progress in both hardware and software development, including, for instance, Google's recent experimental demonstration of 'quantum supremacy' on the task of sampling from random quantum circuits[6], a number of significant knowledge gaps and challenges remain. To surpass classical computers, quantum computer architectures will need to improve numbers of qubits, improve connectivity between qubits and reduce error rates both for operations and for storage, as well as expand algorithmic development into all areas where classical computing faces inherent bottlenecks. These challenges are all significant and are partially conflicting; indeed the central experimental QC challenge is to create quantum systems that are both highly decoupled from unwanted environmental degrees of freedom yet subject to fast and precise control and measurement. While there has been steady experimental progress over the past two decades, it is not easy to predict the rate of future improvements in QC. A recent consensus study on the progress and prospects of quantum computing from the National Academies of Sciences, Engineering and Medicine estimates that to find a private key in a 1024-bit RSA encrypted message using Shor's algorithm requires building a quantum computer that is five orders of magnitude larger and has error rates that are two orders of magnitude better than existing machines[98]. More than 100 academic and government laboratories around the world are working to address these challenges with a variety of hardware solutions[98]. These include ion trap quantum computers with 20-100 qubits that are likely to become available by the early 2020s[98]. Leveraging the power of lithographic technology, super-conducting quantum computers hold great promise, and 5-, 16- and 20-qubit machines are currently available to users via the web. Other promising approaches include developing quantum computers based on photonic, neutral atom and semiconductor qubits[98].

As mentioned above, a number of algorithmic quantum speedups depend on qRAM, but there is no practical implementation of this technology. In fact, this reliance on qRAM, in part, stems from attempts to arrive at quantum algorithms that are essentially quantum versions of classical algorithms. An alternative approach is to design intrinsically quantum algorithms which take advantage of quantum features such as interference. This alternative approach offers the additional benefit that small scale versions of problems are readily implementable on existing hardware. Indeed, recent advances in so-called "near-term" quantum machine learning algorithm development exploit the exponentially large quantum state space to estimate kernel functions[97,99] as well as the natural ability of quantum computers to execute kernel-based classification[100,101]. Generalizations of these algorithms for genomics applications hold great promise and will allow assessment of the current capabilities of publicly available quantum computers[66]. Given the potential of quantum computers to efficiently explore a vast state space, the natural applications to neuroscience problems are largely associated with optimization and machine learning as detailed above. However, yet another path is to identify computational problems that can be naturally cast into a quantum framework. For example, the minimum free energy among all possible protein folds is an important problem with an exponentially large search space and thus a compelling target for quantum algorithm development. Another natural set of problems are those associated with quantum biology – the study of chemical processes including formation of excited electron states within molecules (e.g., proteins) in living cells, and their functional effects[102]. These processes are inherently quantum mechanical and may involve an exponentially vast set of excitation states, which can only be efficiently modeled by applying transformations to an exponentially large state space afforded by a quantum computer. It is unclear whether such processes can be relevant to higher-levels of brain function (and consciousness[103,104]); the algorithms used by the brain at Marr's algorithmic/representational level may be necessarily classical[95], although the advent of quantum machine learning means that increasingly this need not be the case for artificial agents.

While a cautious albeit optimistic estimation associated with steady progress of quantum hardware development (e.g., applying Moore's law) puts the availability of sufficiently powerful, universal quantum computers years in the future, sudden, orders-of-magnitude breakthroughs in resolution, noise reduction, etc. are not unprecedented in experimental physics. Such unforeseen breakthroughs would unleash the power of quantum computing to address pressing computational challenges in biology.